# The "Wow! signal" of the terrestrial genetic code

Vladimir I. *sh*Cherbak[a] and Maxim A. Makukov[b*]

[a]Department of Mathematics, al-Farabi Kazakh National University, Almaty, Republic of Kazakhstan
e-mail: Vladimir.shCherbak@kaznu.kz

[b]Fesenkov Astrophysical Institute, Almaty, Republic of Kazakhstan
e-mail: makukov@gmail.com, makukov@aphi.kz

**ABSTRACT**

It has been repeatedly proposed to expand the scope for SETI, and one of the suggested alternatives to radio is the biological media. Genomic DNA is already used on Earth to store non-biological information. Though smaller in capacity, but stronger in noise immunity is the genetic code. The code is a flexible mapping between codons and amino acids, and this flexibility allows modifying the code artificially. But once fixed, the code might stay unchanged over cosmological timescales; in fact, it is the most durable construct known. Therefore it represents an exceptionally reliable storage for an intelligent signature, if that conforms to biological and thermodynamic requirements. As the actual scenario for the origin of terrestrial life is far from being settled, the proposal that it might have been seeded intentionally cannot be ruled out. A statistically strong intelligent-like "signal" in the genetic code is then a testable consequence of such scenario. Here we show that the terrestrial code displays a thorough precision-type orderliness matching the criteria to be considered an informational signal. Simple arrangements of the code reveal an ensemble of arithmetical and ideographical patterns of the same symbolic language. Accurate and systematic, these underlying patterns appear as a product of precision logic and nontrivial computing rather than of stochastic processes (the null hypothesis that they are due to chance coupled with presumable evolutionary pathways is rejected with $P$-value $< 10^{-13}$). The patterns are profound to the extent that the code mapping itself is uniquely deduced from their algebraic representation. The signal displays readily recognizable hallmarks of artificiality, among which are the symbol of zero, the privileged decimal syntax and semantical symmetries. Besides, extraction of the signal involves logically straightforward but abstract operations, making the patterns essentially irreducible to any natural origin. Plausible way of embedding the signal into the code and possible interpretation of its content are discussed. Overall, while the code is nearly optimized biologically, its limited capacity is used extremely efficiently to store non-biological information.

**Introduction**

Recent biotech achievements make it possible to employ genomic DNA as data storage more durable than any media currently used (Bancroft *et al.*, 2001; Yachie *et al.*, 2008; Ailenberg & Rotstein, 2009). Perhaps the most direct application for that was proposed even before the advent of synthetic biology. Considering alternative informational channels for SETI, Marx (1979) noted that genomes of living cells may provide a good instance for that. He also noted that even more durable is the genetic code. Exposed to strong negative selection, the code stays unchanged for billions of years, except for rare cases of minor variations (Knight *et al.*, 2001) and context-dependent expansions (Yuan *et al.*, 2010). And yet, the mapping between codons and amino acids is malleable, as they interact via modifiable molecules of tRNAs and aminoacyl-tRNA synthetases (Giegé *et al.*, 1998; Ibba & Söll, 2000; see also *Appendix A*). This ability to reassign codons, thought to underlie the evolution of the code to multilevel optimization (Bollenbach *et al.*, 2007), also allows to modify the code artificially (McClain & Foss, 1988; Budisa, 2006; Chin, 2012). It is possible, at least in principle, to arrange a mapping that both conforms to functional requirements and harbors a small message or a signature, allowed by 384 bits of informational capacity of the code. Once genome is appropriately rewritten (Gibson *et al.*, 2010), the new code with a signature will stay frozen in the cell and its progeny, which might then be delivered through space and time to putative recipients. Being energy-efficient (Rose & Wright, 2004) and self-replicating, the biological channel is also free from problems peculiar to radio signals: there is no need to rely on time of arrival, frequency and direction. Thus, due to these restrictions the origin of the famous "Wow!" signal received in 1977 remains uncertain (Ehman, 2011). The biological channel has been given serious considerations for its merits in SETI, though with the focus on genomes (Yokoo & Oshima, 1979; Freitas, 1983; Nakamura, 1986; Davies, 2010; Davies, 2012).

Meanwhile, it has been proposed to secure terrestrial life by seeding exoplanets with living cells (Mautner, 2000; Tepfer, 2008), and that seems to be a matter of time. The biological channel suggests itself in this enterprise. To avoid anthropocentric bias, it might be admitted that terrestrial life is not the starting point in the series of cosmic colonization (Crick & Orgel, 1973; Crick, 1981). If so, it is natural to expect a statistically strong intelligent-like "signal" in the terrestrial genetic code (Marx, 1979). Such possibility is incited further by the fact that how the code came to be apparently non-random and nearly optimized still remains disputable and highly speculative (for reviews on traditional models of the code evolution see Knight *et al.*, 1999; Gusev & Schulze-Makuch, 2004; Di Giulio, 2005; Koonin & Novozhilov, 2009).

The only way to extract a signal, if any, from the code is to arrange its elements – codons, amino acids and syntactic signs – by their parameters using some straightforward logic. These arrangements are then analyzed for patterns or grammar-like structures of some sort. The choice of arrangements and parameters should exclude arbitrariness. For example, only those parameters should be considered which do not depend on systems of physical units. However, even in this case *a priori* it is unknown exactly what kind of patterns one might expect. So there is a risk of false positives, as with a data set like the genetic code it is easy to find various patterns of one kind or another.

Nonetheless, the task might be somewhat alleviated. First, it is possible to predict some general aspects of a putative signal and its "language", especially if one takes advantage of active SETI experience. For example, it is generally accepted that numerical language of arithmetic is the same for the entire universe (Freudenthal, 1960; Minsky, 1985). Besides, symbols and grammar of this language, such as positional numeral systems with zero conception, are hallmarks of intelligence. Thus, interstellar messages sent from the Earth usually began with natural sequence of numbers in binary or decimal notation. To reinforce the artificiality, a symbol of zero was placed in the abstract position preceding the sequence. Those messages also included symbols of arithmetical operations, Egyptian triangle, DNA and other notions of human consciousness (Sagan *et al.*, 1972; The Staff at the NAIC, 1975; Sagan *et al.*, 1978; Dumas & Dutil, 2004).

Second, to minimize the risk of false positives one can impose requirements as restrictive as possible on a putative signal. For example, it is reasonable to expect that a genuinely intelligent message would represent not just a collection of patterns of various sorts, but patterns of the same "linguistic style". In this case, if a potential pattern is noticed, further search might be narrowed down to the same sort of patterns. Another stringent requirement might be that patterns should involve each element of the code in each arrangement, whereas the entire signal should occupy most, if not all, of the code's informational capacity. By and large, given the nature of the task, specifics of the strategy are defined en route.

Following these lines, we show that the terrestrial code harbors an ensemble of precision-type patterns matching the requirements mentioned above. Simple systematization of the code reveals a strong informational signal comprising arithmetical and ideographical components. Remarkably, independent patterns of the signal are all expressed in a common symbolic language. We show that the signal is statistically significant, employs informational capacity of the code entirely, and is untraceable to natural origin. The models of emergence of primordial life with original signal-free genetic code are beyond the scope of this paper; whatever it was, the earlier state of the code is erased by palimpsest of the signal.

**Background**

Should there be a signal in the code, it would likely have manifested itself someway during the half-century history of traditional analysis of the code organization. So it is of use to summarize briefly what has been learned about that up to date. Also, for the sake of simplicity in data presentation, we will mention in advance some *a posteriori* information concerning the signal to be described, with fuller discussion in due course. We suggest to a reader unfamiliar with molecular mechanisms behind the genetic code first to refer to *Appendix A*, where it is also explained why the code is amenable to intentional "modulation" (to use the language of radio-oriented SETI) and, at the same time, is highly protected from casual "modulation" (has strong noise immunity).

***The code at a glance.*** As soon as the genetic code was biochemically cracked (Nirenberg *et al.*, 1965), its non-random structure became evident (Woese, 1965; Crick, 1968). The most obvious pattern that emerged in the code was its regular redundancy. The code comprises 16 codon families beginning with the same pair of bases, and these families generally consist of either one or two equal series of codons mapped to one amino acid or to *Stop* (Fig. 1a). In effect, the standard code is nearly symmetric in redundancy. There are only two families split unequally: those beginning with TG and AT. The minimum action to restore the symmetry is to match TG-family against AT-family by reassigning TGA from *Stop* to cysteine. Incidentally, this symmetrized version is not just a theoretical guess but is also found in nature as the nuclear code of euplotid ciliates (Meyer *et al.*, 1991). While the standard code stores the arithmetical component of the signal, the symmetrical euplotid version keeps the ideographical one (the interrelation between these two code versions is discussed later). Regular redundancy leads also to the block structure of the genetic code. This makes it possible to depict the code in a contracted form, where each amino acid corresponds to a single block, or a contracted series (Fig. 1b). The three exceptions are Arg, Leu and Ser, which have one IV-series and one II-series each.

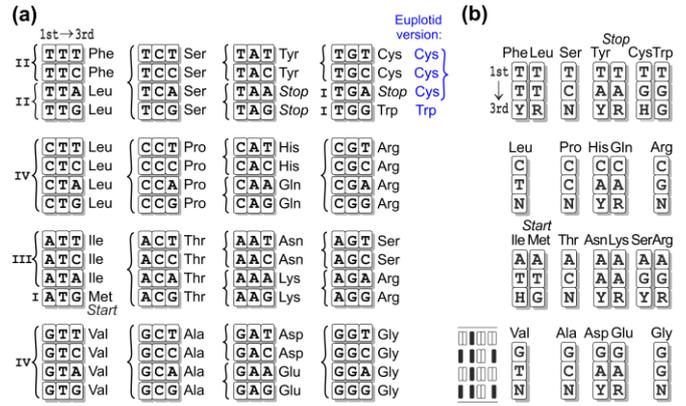

**Fig. 1.** The genetic code. (**a**) Traditional representation of the standard, or universal, code. Codons coding the same amino acid form synonymic series denoted with opening braces. Number of codons in a series defines its redundancy (degeneracy). Whole codon families consist of one series of redundancy IV. Other families are split. Most split families are halved into two series of redundancy II each, one ending with pyrimidines {T, C} and another with purines {A, G}. Three codons in the standard code are not mapped to any amino acid and are used as *Stop* in translation. The *Start* is usually signified by ATG which codes Met. Closing brace shows the only difference between the euplotid and the standard code. (**b**) Contracted representation of the euplotid version. Synonymous full-size codons are replaced by a single contracted series with combined third base. FASTA designations are used: R and Y stand for purines and pyrimidines, respectively, N stands for all four bases and H stands for {T, C, A}. Series are placed vertically for further convenience. The pictogram on the left helps in figures below. Filled elements denote whole families here.

Apart from regular redundancy, a wealth of other features were reported afterwards, among which are robustness to errors (Alff-Steinberger, 1969), correlation between thermostability and redundancy of codon families (Lagerkvist, 1978), non-random distribution of amino acids among codons if judged by their polarity and bulkiness (Jungck, 1978), biosynthetic pathways (Taylor & Coates, 1989), reactivity (Siemion & Stefanowicz, 1992), and even taste (Zhuravlev, 2002). The code was also shown to be effective at handling additional information in DNA (Baisnée *et al.*, 2001; Itzkovitz & Alon, 2007). Apparently, these features are related, if anything, to the direct biological function of the code. There are also a number of abstract approaches to the code, such as those based on topology (Karasev & Stefanov, 2001), information science (Alvager *et al.*, 1989), and number theory (Dragovich, 2012). However, the main focus of these approaches is in constructing theoretical model descriptions of known features in the code, rather than dealing with new ones.

All in all, only two intrinsic regularities, observed early on in the study of the code, might suggest possible relation to a putative signal due to their conspicuous and unambiguous character. They also suggest two dimensionless integer parameters for signal extraction. These are quantity of codons in a series mapped to one amino acid (redundancy) and quantity of nucleons in amino acid molecules. These parameters might be called "ostensive numerals" by analogy with the quantity of radio beeps in *Lingua Cosmica* (Freudenthal, 1960).

***Rumer's bisection.*** Rumer (1966) bisected the code by redundancy – the first "ostensive numeral". There are 8 whole families and 8 split families in the code (Fig. 2a). Rumer found that codons in these families are mapped to each other in a one-to-one fashion with a simple relation T↔G, C↔A, now known as Rumer's transformation. There are two more transformations of such type: T↔C, A↔G and T↔A, C↔G. They also appear in Rumer's bisection and each makes half of what Rumer's transformation makes alone.

Arbitrary bisection of the code has small chances to produce a transformation, and still less – their ordered set (see *Appendix B*). Rumer's finding was rediscovered by Danckwerts & Neubert (1975), who also noted that this set might be described with a structure known in mathematics as the Klein-4 group. That triggered a series of yet other models involving group theory to describe the code (Bertman & Jungck, 1979; Hornos & Hornos, 1993; Bashford et al., 1998), which, admittedly, did not gain decisive insights. Meanwhile, in traditional theories of the code evolution this feature was ignored altogether, though it was repeatedly rediscovered again (e.g., see Wilhelm & Nikolajewa, 2004). Noteworthy, this regularity – which turns out to be a small portion of the signal – was first noticed immediately after codon assignments were elucidated. Together with the fact of rediscoveries, this speaks for the anticryptographic nature of the signal inside the code.

***Amino acid nucleons.*** Hasegawa & Miyata (1980) arranged amino acids in order of increasing nucleon number – the second "ostensive numeral" which, unlike other amino acid properties, does not rely on arbitrarily chosen system of units. Such arrangement reveals a rough anticorrelation: the greater the redundancy the smaller the nucleon number (Fig. 2b). This promoted speculations that prevailing small amino acids occupied the series of higher redundancy during the code evolution. As shown below, this anticorrelation is a derivative of the signal. Moreover, exactly this observation suggests simple systematization for both "ostensive numerals": monotonous arraying of nucleon and redundancy numbers in opposite directions.

On the whole, Hasegawa and Miyata dealt with amino acids whereas Rumer dealt with codons. Combined, these approaches yield assignments between codons and amino acid nucleon numbers convenient for systematization. *Stop*-codons code for no amino acid; therefore, to include them into the systematization, they are assigned a zero nucleon number.

***The activation key.*** All arithmetical patterns considered further appear with the differentiation between blocks and chains in all 20 amino acids and with the subsequent transfer of one nucleon from side chain to block in proline (Fig. 2b). Proline is the only exception from the general structure of amino acids: it holds its side chain with two bonds and has one hydrogen less in its block. The mentioned transfer in proline "standardizes" its block nucleon number to 73 + 1 and reduces its chain nucleons to 42 – 1. In itself, the distinction between blocks and chains is purely formal: there is no stage in protein synthesis where amino acid side chains are detached from standard blocks. Therefore, there is no any natural reason for nucleon transfer in proline; it can be simulated only in the mind of a recipient to achieve the array of amino acids with uniform structure. Such nucleon transfer thus appears artificial. However, exactly this seems to be its destination: it protects the patterns from any natural explanation. Minimizing the chances for appealing to natural origin is a distinct concern in messaging of such kind, and this problem seems to be solved perfectly for the signal in the genetic code. Applied systematically without exceptions, the artificial transfer in proline enables holistic and arithmetically precise order in the code. Thus, it acts as an "activation key". While nature deals with the actual proline which does not produce the signal in the code, an intelligent recipient easily finds the key and reads messages in arithmetical language (see also *Discussion*).

***Decimalism.*** The arithmetical patterns to be described hold true in any numeral system. However, as it turned out, expressed in positional decimal system, they all acquire conspicuously distinctive notation. Therefore, here we briefly provide some relevant information.

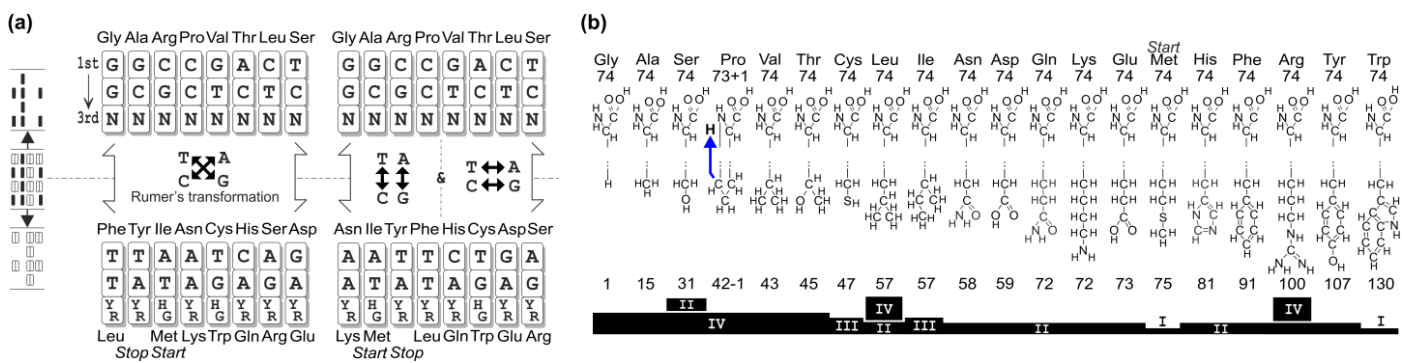

**Fig. 2.** Preceding observations. (**a**) Rumer's bisection. Whole families are opposed to split ones, thereby bisecting the code. Codons in opposed families are mapped to each other with the ordered set of Rumer's transformation and two half-transformations. Transformation of third bases is trivial as they are the same in any family; therefore contracted representation is adequate to show this regularity. The regularity is valid both for the standard and the euplotid (shown here) version. (**b**) Categorization of amino acids by nucleon numbers. Free molecules unmodified by cytoplasmic environment are shown. Each of them is formed of the standard block and a side chain. Blocks are identical in all amino acids except proline. Chains are unique for each amino acid. Numbers of nucleons, i.e. protons and neutrons, are shown for both blocks and chains. To avoid ambiguity, it is judicious to consider only most common and stable isotopes: $^{1}$H, $^{12}$C, $^{14}$N, $^{16}$O, $^{32}$S. The bar at the bottom shows the redundancy of amino acids in the code. Cross-cut bonds symbolize the distinction between standard blocks and unique side chains of amino acids. The arrow in proline denotes hereafter the "activation key" (see text).

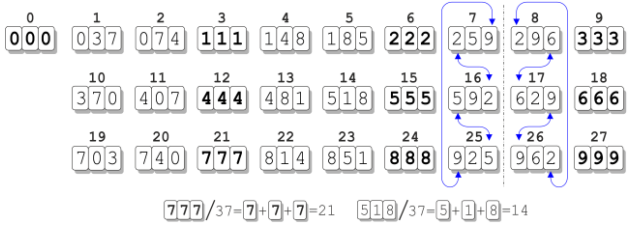

**Fig. 3.** Digital symmetry of decimals divisible by *037*. Leading zero emphasizes its equal participation in the symmetry. All three-digit decimals with identical digits 111, …, 999 are divisible by *037*. The sum of three identical digits gives the quotient of the number divided by *037*. Analogous sum for numbers with unique digits gives the central quotient in the column. Digits in these numbers are interconnected with cyclic permutations that are mirror symmetrical in neighbor columns. Addition instead of division provides an efficient way to perform checksums (see *Appendix C*). The scheme extends to decimals with more than three digits, if they are represented as $a + 999 \times n$, where $n$ is the quotient of the number divided by 999 and $a$ is the remainder, to which the same symmetry then applies (for three-digit decimals $n = 0$). Numbers divisible by *037* and larger than 999 will be shown in this way.

Nature is indifferent to numerical languages contrived by intelligence to represent quantities, including zero. A privileged numeral system is therefore a reliable sign of artificiality. Intentionally embedded in an object, a privileged system might then demonstrate itself through distinctive notation to any recipient dealing with enumerable elements of that object. For example, digital symmetries of numbers divisible by prime *037* exist only in the positional decimal system with zero conception (Fig. 3). Thus, distinctive decimals 111, 222 and 333 look ordinarily 157, 336 and 515 in the octal system. This notational feature was marked by Pacioli (1508) soon after the decimal system came to Europe. Analogous three-digit feature exists in some other systems, including the quaternary one (see *Appendix C*).

### Results

The overall structure of the signal is shown in Fig. 4, which might be used as guidance in further description. The signal is composed of arithmetical and ideographical patterns, where arithmetical units are represented by amino acid nucleons, whereas codon bases serve as ideographical entities. The patterns of the signal are displayed in distinct logical arrangements of the code, thereby increasing both the informational content of the signal and its statistical significance. Remarkably, all of the patterns bare the same general style reflected in Fig. 4 with identical symbols in each signal component (represented by boxes). Namely, distinct logical arrangements of the code and activation key produce exact equalities of nucleon sums, which furthermore display decimalism and are accompanied by Rumer's and/or half-transformations. One of these arrangements furthermore leads to ideography and semantical symmetries. All elements of the code – 64 codons, 20 amino acids, *Start* and *Stop* syntactic signs – are involved in each arrangement.

Unlike radio signals which unfold in time and thus have sequential structure, the signal in the genetic code has no entry point, similar to the pictorial message of Pioneer plaques (Sagan *et al.*, 1972). However, instead of providing pictograms the signal in the genetic code provides patterns that do not depend on visual symbols chosen to represent them (be it symbols for nucleotide bases or for the notation of "ostensive numerals"). These patterns make up the organic whole, so there is no unique order in presenting them. We will begin with arithmetical component and then move on to ideography.

*The arithmetical component*

**Full-size standard code.** One logically plain arrangement of the code was proposed by George Gamow in his attempt to guess the coding assignments theoretically before the code was cracked *in vitro* (see Hayes, 1998). One of his models, though it did not predict the actual mapping correctly, coincided remarkably with one of the signal components. Gamow arranged codons according to their composition, since 20 combinations of four bases taken three at a time could account for 20 amino acids (Gamow & Yčas, 1955). Bringing nucleon numbers, activation key and few "freezing" conditions into this arrangement reveals total nucleon balancing ornate with decimal syntax.

Codons with identical and unique bases comprise two smaller sets (Fig. 5a). Halved, both sets show the balance of side chains with $703 = 037 \times 019$ nucleons in each half as well as the balance of whole molecules with $1665 = 666 + 999 \times 1$ nucleons. Importantly, the halving is not arbitrary. Codons are opposed by Rumer's transformation along with the half-transformation T↔C, A↔G in the first set and T↔A, C↔G in the second set.

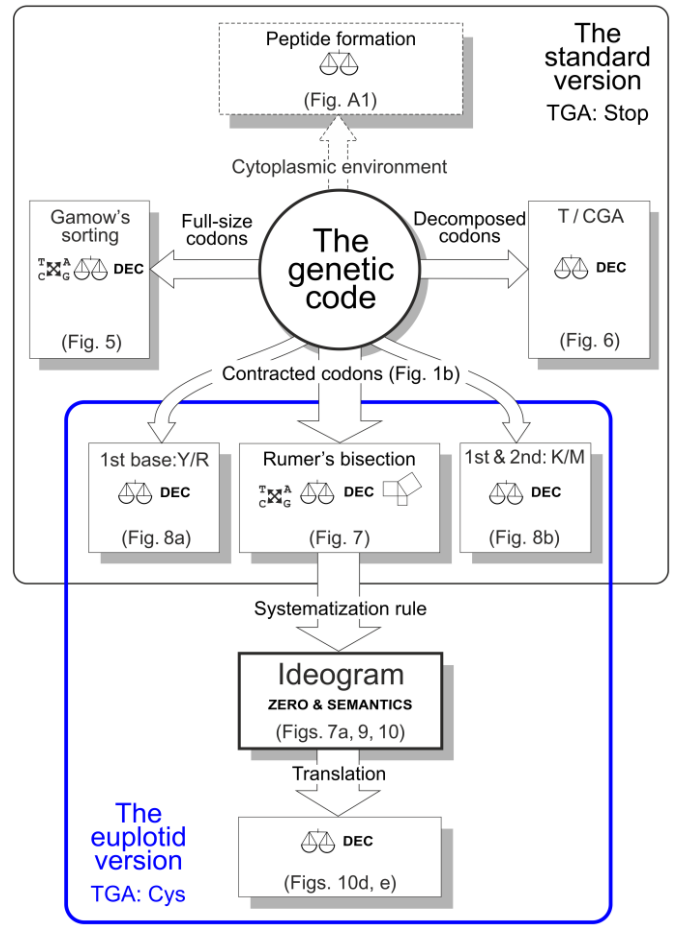

**Fig. 4.** The structure of the signal. All details are discussed sequentially in the text. The image of scales represents precise nucleon equalities. DEC stands for distinctive decimal notation of nucleon sums. The dotted box denotes the cytoplasmic balance (see *Appendix D*), the only pattern maintained by actual proline and cellular milieu. All other patterns are enabled by the "activation key" and are valid for free amino acids. K stands for {T, G}, M stands for {A, C}. Though all three types of transformations act in the patterns, only Rumer's transformation is indicated for simplicity.

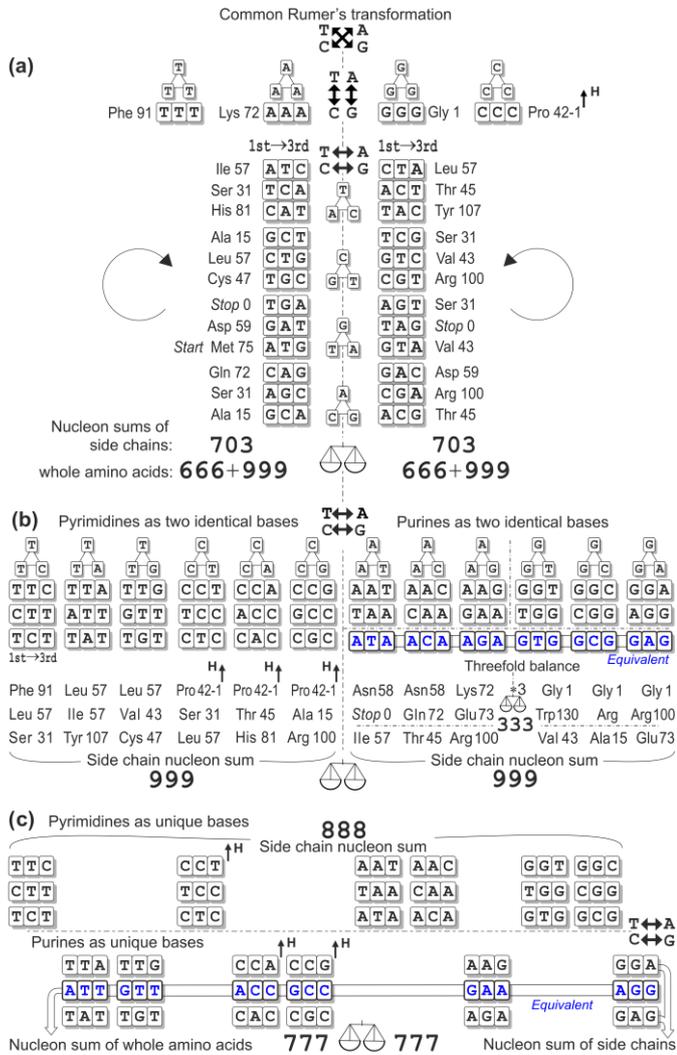

tical bases type. Though not balanced, these halves again show distinctive decimal syntax with 888 and 1110 = 111 + 999×1 nucleons. Decimalism of one of these sums is algebraically dependent, as from the previous case (Fig. 5b) the sum of the whole set is known to be divisible by *037*; if a part of this set is decimally distinctive, the other one will be such automatically. Notably, an independent pattern nonetheless stands out here. Namely, a part of the previous threefold balance has an equivalent in one half here, where the same amino acids are represented by synonymous codons (Fig. 5b and c). Whole molecules of this equivalent – 333 side chain and 444 standard block nucleons – are balanced with 777 chain nucleons in the rest of the subset.

Note that all those distinctive notations of nucleon sums appear only in positional decimal system. Positional notation is so customary in our culture that most of its users hardly remember a fairly complex rule behind it that encodes numbers as $a_{n-1} \times q^{n-1} + \ldots + a_1 \times q^1 + a_0 \times q^0$, where $q = 10$ in case of the decimal system, $n$ is the quantity of digits in notation, and $a_i$ – digits 0-9 that are left in the final notation.

***Decomposed standard code.*** Another arrangement of the code is brought about by decomposition of its 64 full-size codons. This yields 192 separate bases and reveals a pattern of the same type as in full-size format. Identical bases make up four sets of 48 bases in each. Each base retains the amino acid or *Stop* of its original codon (Fig. 6a). Thus, the four sets get their individual chain and block nucleon sums.

In total, there are 222 + 999×10 side chain nucleons in the decomposed code – obviously, thrice as much as the total sum in the previous full-size case (with the activation key still applied). Only one combination of the four sets displays distinctive decimalism of side chain nucleon sums. These are 666 + 999×2 nucleons in the T-set and 555 + 999×7 nucleons in the joint CGA-set (Fig. 6b). Meanwhile, there are exactly 222 + 999×10 block nucleons in the CGA-set (note that the sets have unequal block sums due to different accumulation of *Stops*). Thus, while chain nucleons are outnumbered by block nucleons overall the code, they are neatly balanced with their CGA-part.

***Contracted code and the systematization rule.*** In a sense, contraction of codon series (see Fig. 1b) is an operation logically opposite to decomposition. Besides displaying new arithmetical patterns, contracted code also reveals ideographical component

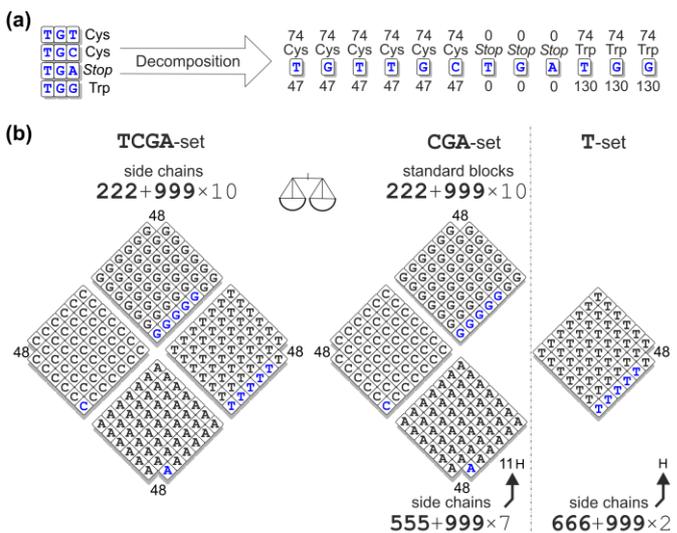

**Fig. 5.** Gamow's sorting of codons according to their nucleotide base composition. Base combinations (shown on triangular frames) produce three sets: 4 codons with three identical bases, 24 codons with unique bases and 36 codons with two identical bases. (**a**) The first and the second sets halved by vertical axis with Rumer's and half-transformations along with *Spin → Antispin* transformation denoted with circular arrows. Applied to triangular frames, these arrows define the sequence of bases in codons. Note that while any block sum (with the activation key applied) is divisible by *037* as each block has 74 = 2×*037* nucleons, chain sums are not restricted in this way. (**b**) The third set halved according to whether identical bases are purines or pyrimidines. (**c**) The third set halved with horizontal axis according to whether unique bases are purines or pyrimidines.

The *Spin → Antispin* transformation does not affect the first set but finally freezes elements of the second one. There is only one degree of freedom left since there are no reversible transformations that might connect both sets, so one of them is free to swap around the axis. The balance appears in one of the two alternative states.

The third set includes codons with two identical bases. When halved according to whether they are purines or pyrimidines, regardless of the unique base type, this set shows the balance 999 = 999 of side chains (Fig. 5b). Besides, such halving keeps Rumer's and one of the half-transformations again in place. In its turn, the right half of the set is threefold balanced. Codons with adenine side by side, guanine side by side and palindromic codons make up three equal parts with 333 nucleons each.

In Fig. 5c the same set is halved according to whether unique bases are purines or pyrimidines, this time regardless of the iden-of the signal. The systematization rule leading to the ideography combines findings of Rumer (1966) and of Hasegawa & Miyata (1980) and is symmetric in its nature (*sh*Cherbak, 1993). Contracted series are sorted into four sets according to their redundancy; within those sets they are aligned side-by-side in order of monotonously changing (e.g., increasing) nucleon number. The sets themselves are then arranged in antisymmetrical fashion (e.g., in order of decreasing redundancy number). *Stop*-series is placed at the beginning of its set representing zero in its special position. Finally, Rumer's bisection opposes the IV-set to III, II, I sets. The resulting arrangement is shown in Fig. 7 for the euplotid code, with ideography of codon bases (see next section) in Fig. 7a and arithmetical patterns of amino acids (shared by both code versions) in Fig. 7b.

**Fig. 6.** The decomposed standard code. (**a**) Decomposition shown for one family of codons. Three T-bases contribute three Cys molecules into T-set; one A-base contributes one *Stop* to A-set and so on for the entire code. (**b**) Identical bases are sorted into four sets regardless of their position in codons. The sets are shown twice for convenience.

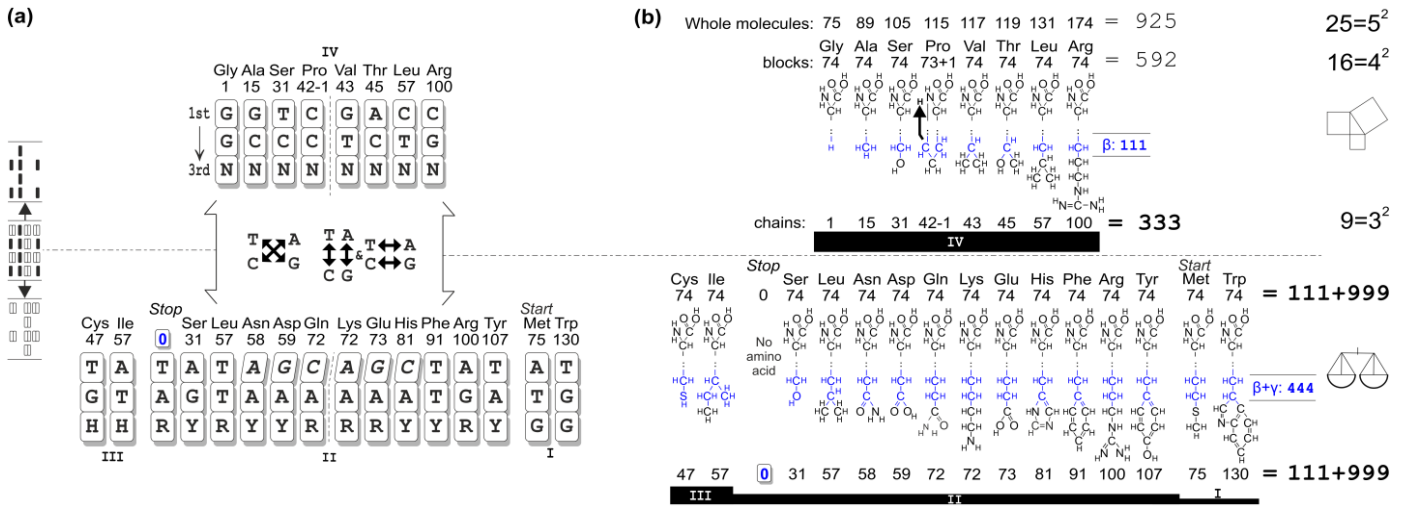

**Fig. 7.** The contracted euplotid code with the systematization rule applied (compare with Fig. 2). (**a**) The resulting arrangement of contracted codon series forming the ideogram. Side-by-side alignment of vertical series produces three horizontal strings of peer-positioned bases. Gln and Lys have the same nucleon number; ambiguity in their positioning is eliminated by the symmetries considered further. (**b**) The arithmetical background of the ideogram (valid for the standard version as well, as it contributes another zero to the III, II, I set). For β and γ side chain levels see *Discussion*.

A new balance is found in the joint III, II, I set. Side chain nucleons of all its amino acids are equalized with their standard blocks: $111 + 999 \times 1 = 111 + 999 \times 1$ (Fig. 7b). This pattern manifests as the anticorrelation mentioned by Hasegawa & Miyata (1980). Chain nucleon sum of all series in the code is less than the sum of all blocks. Only a subset of series coding mainly bigger amino acids may equalize its own blocks. Exactly this happens in the joint III, II, I set. As a consequence, smaller amino acids are left in the set of redundancy IV.

Meanwhile, there are 333 chain and 592 block nucleons and $333 + 592 = 925$ nucleons of whole molecules in the IV-set. With *037* cancelled out, this leads to $3^2 + 4^2 = 5^2$ – numerical representation of the Egyptian triangle, possibly as a symbol of two-dimensional space. Incidentally, codon series in the ideogram (Fig. 7a) are arranged in the plane rather than linearly in a genomic fashion.

Rumer's bisection is based on redundancy and thus makes use of third positions in codon series. Divisions of the contracted code based on first and center positions also reveal similar patterns (Fig. 8). Another arithmetical phenomenon presumably related to the signal – the cytoplasmic balance – is described in *Appendix D*.

Thus, the standard code reveals same-style and yet algebraically independent patterns simultaneously in decomposed, full-size, and contracted representations (see Fig. 4). It is a highly nontrivial algebraic task to find the solution that maps amino acids and syntactic signs to codons in a similar fashion. Normally this would require considerable computational power.

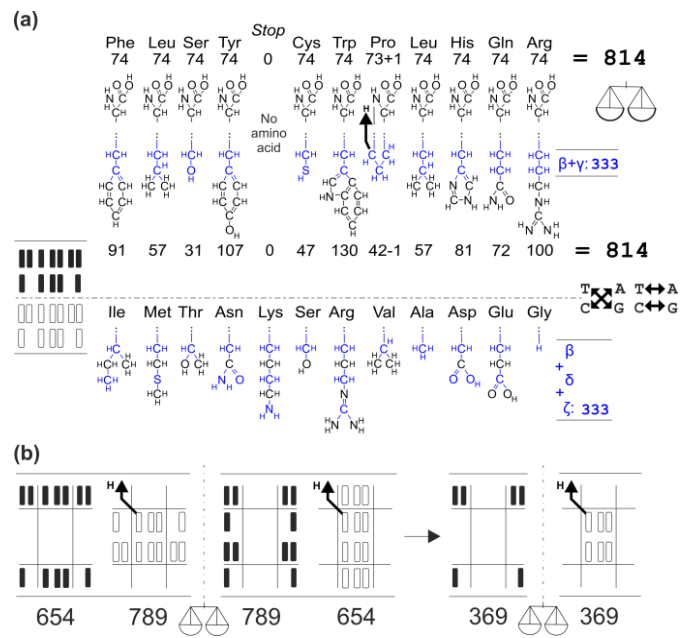

**Fig. 8.** Additional arithmetical patterns of the contracted code (shared by both code versions). (**a**) The code is divided according to whether first bases are purines or pyrimidines. This gives two sets with equal numbers of series. The halve with pyrimidines in first positions reveals a new balance of chains and blocks analogous to that in Fig. 7b. Another halve is algebraically dependent except the decimal sum of its β, δ, ζ levels, see *Discussion*. (**b**) The code is divided according to whether first bases are K or M (left) or whether central bases are K or M (center). Both divisions produce halves with identical chain nucleon sums. As algebraic consequence of these divisions, series with K in first and central positions and series with M in first and central positions are chain-balanced (right). Each of the three divisions is accompanied by half-transformations and, remarkably, also produces equal numbers of series in each half. This pattern is the only one that shows no divisibility by *037*. However, all three numbers – 654, 789 and 369 – are again specific in decimal notation where digits in each of them appear as arithmetic progressions.

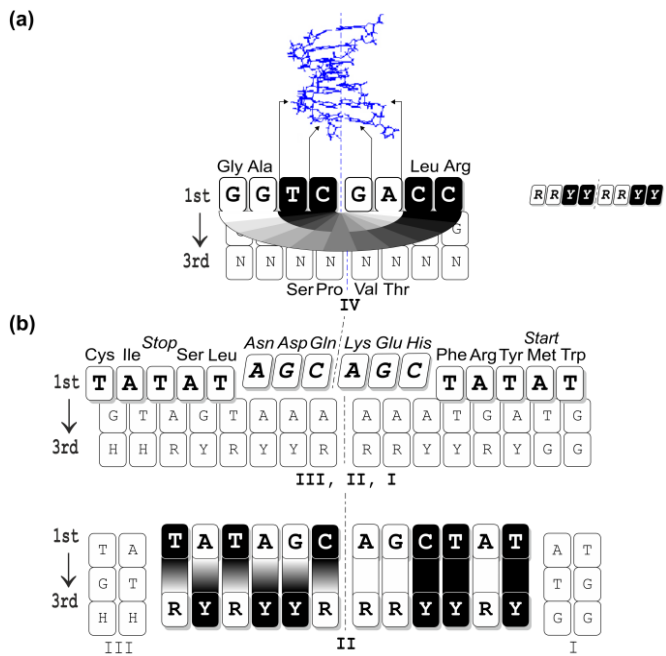

**Fig. 9.** Patterns of the short (**a**) and the long (**b**) upper strings. The strings are arranged with the same set of symmetries: *mirror* symmetry (denoted with the central vertical axis), *translation* symmetry (denoted with italicized letters and skewed frames) and purine ↔ pyrimidine *inversion* (denoted with color gradient, where black and white stand for pyrimidines and purines, respectively). The image of DNA at the top illustrates possible interpretation of the short string (see *Discussion*).

*The ideographical component*

***Upper strings.*** We refer to the product of systematization in Fig. 7a as the ideogram. The ideogram of the genetic code is based on symmetries of its strings (*sh*Cherbak, 1988). The strings are read across contracted series.

The upper short string demonstrates *mirror*, *translation* and *inversion* symmetries (Fig. 9a). Its bases are invariant under combined operation of the *mirror* symmetry and *inversion* of the type base → complementary base. A minimum pattern of the *translation* symmetry is represented by *RRYY* quadruplet.

The same three symmetries arrange the long upper string (Fig. 9b). The pair of flanking TATAT sequences is *mirror* symmetrical. The pair of central *AGC* codons forms a minimum pattern of the *translation* symmetry. First and third bases in the set of redundancy II are interconnected in an axisymmetric manner with purine ↔ pyrimidine *inversion* and its opposite operation – the unit transformation producing no exchange.

***Center strings.*** Placed coaxially, the short and the long center strings appear interconnected with purine ↔ pyrimidine *inversion* (Fig. 10a). Both strings exhibit purine-pyrimidine *mirror* symmetry. The long string keeps the mirror symmetry even for ordinary bases.

Codons of the short string CCC and TCT break the mirror symmetry of ordinary bases, but they share a palindromic feature, i.e. direction of reading invariance. This feature restores the mirror symmetry, this time of the *semantical* type (Fig. 10b). As in the previous case, two center strings are expected to share the same set of symmetries. Therefore, the semantical symmetry of palindromic codons flanked by G-bases may indicate a similar feature in the long string. Indeed, semantical symmetry is found there in the triplet reading frame starting after flanking G-base (Fig. 10c). This reading frame is remarkable with the regular arrangement of all syntactic signs of the euplotid code – both *Stop*-codons and the *Start*-codon repeated twice. The reading frame displays the *semantical mirror* symmetry of antonyms with homogeneous AAA-codon in the center.

The codons of this reading frame are purely abstract symbols, given that they are read across contracted series. However, they are regularly crossed with the same codons in the ideogram, thereby reinforcing the semantical symmetry and making the current frame unique (Fig 10c). Besides, direction of reading now becomes distinguished since such "crossword" disappears if read in opposite way, though the palindrome itself remains the same.

Remarkably, the triplet string in Fig. 10c is written with the code symbols within the code itself. This implies that the signal-harboring mapping had to be projected preliminarily (see *Discussion*). Besides, translation of this string with the code itself reveals the balance 222 = 222 of chains and blocks (Fig. 10d). Additional palindrome in the frame shifted by one position (Fig. 10e) reproduces the chain sum of 222, confirming that the ideogram is properly "tuned in" to the euplotid version: TGA stands for Cys here, not for *Stop* of the standard code.

## Discussion

***Artificiality.*** To be considered unambiguously as an intelligent signal, any patterns in the code must satisfy the following two criteria: (1) they must be highly significant statistically and (2) not only must they possess intelligent-like features (Elliott, 2010), but they should be inconsistent in principle with any natural process, be it Darwinian (Freeland, 2002) or Lamarckian (Vetsigian et al., 2006) evolution, driven by amino acid biosynthesis (Wong, 2005), genomic changes (Sella & Ardell, 2006), affinities between (anti)codons and amino acids (Yarus et al., 2009), selection for the increased diversity of proteins (Higgs, 2009), energetics of codon-anticodon interactions (Klump, 2006; Travers, 2006), or various pre-translational mechanisms (Wolf & Koonin, 2007; Rodin et al., 2011).

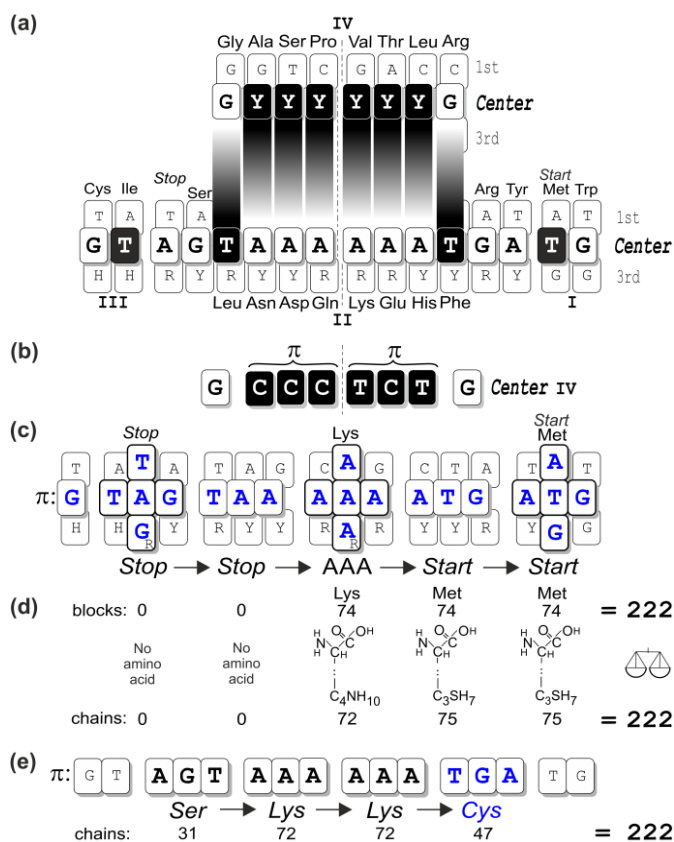

**Fig. 10.** Patterns of the short (**a**, **b**) and the long (**a**, **c**, **d**, **e**) center strings. Both strings are arranged with purine-pyrimidine *mirror* symmetry, purine ↔ pyrimidine *inversion* and *semantical* symmetry. The first two are denoted in the same way as in Fig. 9, π denotes palindrome.

The statistical test for the first criterion is outlined in *Appendix B*, showing that the described patterns are highly significant. The second criterion might seem unverifiable, as the patterns may result from a natural process currently unknown. But this criterion is equivalent to asking if it is possible at all to embed informational patterns into the code so that they could be unequivocally interpreted as an intelligent signature. The answer seems to be yes, and one way to do so is to make patterns virtual, not actual. Exactly that is observed in the genetic code. Strict balances and their decimal syntax appear only with the application of the "activation key". Physically, there are no strict balances in the code (e.g., in Fig. 5b one would have 1002 ≠ 999 instead of 999 = 999). Artificial transfer of a nucleon in proline turns the arithmetical patterns on and thereby makes them virtual. This is also the reason why we interpret distinctive notation as an indication of decimalism, rather than as a physical requirement (yet unknown) for nucleon sums to be multiples of *037*: in general, physically there is no such multiplicity in the code. In its turn, notationally preferred numeral system is by itself a strong sign of artificiality. It is also worth noting that all three-digit decimals – 111, 222, 333, 444, 555, 666, 777, 888, 999 (as well as zero, see below) – are represented at least once in the signal, which also looks like an intentional feature.

However, it might be hypothesized that amino acid mass is driven by selection (or any other natural process) to be distributed in the code in a particular way leading to approximate mass equalities and thus making strict nucleon balances just a likely epiphenomenon. But it is hardly imaginable how a natural process can drive mass distribution in abstract representations of the code where codons are decomposed into bases or contracted by redundancy. Besides, nucleon equalities hold true for free amino acids, and yet in these free molecules side chains and standard blocks had to be treated by that process separately. Furthermore, no natural process can drive mass distribution to produce the balance in Fig. 10d: amino acids and syntactic signs that make up this balance are entirely abstract since they are produced by translation of a string read across codons.

Another way to make patterns irreducible to natural events is to involve semantics, since no natural process is capable of interpreting abstract symbols. It should be noted that notions of symbols and meanings are used sometimes in a natural sense (Eigen & Winkler, 1983), especially in the context of biosemiotics (Barbieri, 2008) and molecular codes (Tlusty, 2010). The genetic code itself is regarded there as a "natural convention" that relates symbols (codons) to their meanings (amino acids). However, these approaches make distinction between organic semantics of molecular codes and interpretive or linguistic semantics peculiar to intelligence (Barbieri, 2008). Exactly the latter type of semantics is revealed in the signal of the genetic code. It is displayed there not only in the symmetry of antonymous syntactic signs (Fig. 10c), but also in the symbol of zero. For genetic molecular machinery there is no zero, there are nucleotide triplets recognized sterically by release factors at the ribosome. Zero – the supreme abstraction of arithmetic – is the interpretive meaning assigned to *Stop*-codons, and its correctness is confirmed by the fact that, being placed in its proper front position, zero maintains all ideogram symmetries. Thus, a trivial summand in balances, zero, however, appears as an *ordinal* number in the ideogram. In other words, besides being an integral part of the decimal system, zero acts also as an individual symbol in the code.

In total, not only the signal itself reveals intelligent-like features – strict nucleon equalities, their distinctive decimal notation, logical transformations accompanying the equalities, the symbol of zero and semantical symmetries, but the very method of its extraction involves abstract operations – consideration of idealized (free and unmodified) molecules, distinction between their blocks and chains, the activation key, contraction and decomposition of codons. We find that taken together all these aspects point at artificial nature of the patterns.

Though the decimal system in the signal might seem a serendipitous coincidence, there are few possible explanations, from 10-digit anatomy as an evolutionary near-optimum for bilateral beings (Dennett, 1996) to the fact that there are conveniently 74 = 2×*037* nucleons in the standard blocks of α-amino acids. Besides, the decimal system shares the triplet digital symmetry with the quaternary one (see *Appendix C*), establishing a link to the "native" language of DNA. After all, some of the messages sent from the Earth included the decimal system as well (Sagan et al., 1978; Dumas & Dutil, 2004), though they were not supposed to be received necessarily by 10-digit extraterrestrials. Whatever the actual reason behind the decimal system in the code, it appears that it was invented outside the Solar System already several billions years ago.

*Two versions of the code.* The nearly symmetric code version with arithmetical patterns acts as the universal standard code. With this code at hand it is intuitively easy to infer the symmetric version with its ideography. Vice versa, if the symmetric version were the universal one, it would be hardly possible to infer the nearly symmetric code with all its arithmetical patterns. Therefore, with the standard version alone it is possible to "receive" both arithmetical and ideographical components of the signal, even if the symmetric version was not found in nature. There are two possible reasons why it is actually found in euplotid ciliates: either originally when Earth was seeded there were both versions of the code with one of them remaining currently in euplotid ciliates, or originally there was only the standard version, and later casual modification in euplotid lineage coincided with the symmetric version.

What concerns other known rare versions of the code, they seem neither to have profound pattern ensembles, nor to be easily inferable from the standard code. As commonly accepted, they represent later casual deviations of the standard code caused by ambiguous intermediates or codon captures (Moura et al., 2010).

*Embedding the signal.* To obtain a code with a signature one might search through all variant mappings and select the "most interesting" one. However, this method is unpractical (at least with the present-day terrestrial computing facilities), given the astronomically huge number of variant codes. In a more realistic alternative, the pattern ensemble of the signal is projected preliminarily as a system of algebraic expressions which is then solved relatively easily to deduce the mapping of the code. Thus, all described patterns might be represented *post factum* as a system of Diophantine expressions (i.e. equations and inequalities allowing only integer solutions), and analysis of this system shows that it uniquely determines the mapping between codon series and nucleon numbers, including zeros for *Stop*-codons (see *Appendix E*). Though some amino acids have equal nucleon numbers, as the case for Leu and Ile, or Lys and Gln, even they are not interchangeable, as suggested by distinctive notation of nucleon sums in β, γ and other positional levels of side chains in the contracted code (Figs. 7b and 8a). The activation key applies here as well (note that β- and δ-carbons in proline are positionally equivalent). The standard chemical nomenclature of carbon atoms is extended here to denote positions of other nodal atoms. Decimalism in different combinations of levels circumvents algebraic dependence and employs chemical structure of amino acids more efficiently.

These patterns within side chains go even deeper into chemical structure. Some of the canonical amino acids – His, Arg and Trp – might exist in alternative neutral tautomeric forms differing in the position of one hydrogen atom in their side chains (Taniguchi & Hino, 1981; Rak *et al.*, 2001; Li & Hong, 2011). Though some of these tautomers occur very rarely at cytoplasmic pH (as the case for indolenine tautomer of Trp shown in Fig. 7b), all neutral tautomers are legitimate if idealized free molecules are considered, and taking only one of them would introduce arbitrariness. Notably, however, that while one Trp tautomer maintains the patterns in Fig. 7b, another one does the job in Fig. 8a, whereas any neutral tautomer of His and Arg might be taken in both cases without affecting the patterns at all (which is easily checked; to this end, both Arg tautomers are shown in Fig. 8a and both His tautomers are shown in Figs. 7b and 8a).

Importantly, preliminary projecting of a signal admits imposition of functional requirements as extra formal conditions. The terrestrial code is known to be conservative with respect to polar requirement (Freeland & Hurst, 1998), but not to molecular size (Haig & Hurst, 1991). The signal in the code does not involve polar requirement as such, so it might be used in a parallel formal condition to reduce effect of misreadings. However, the signal does involve nucleon numbers which correlate with molecular volume. That interferes with an attempt to make the code conservative with respect to size of amino acids as well.

*Possible interpretation.* Besides having the function of an intelligent signature as such, the signal in the genetic code might also admit sensible interpretations of its content. Without claim to be correct, here we propose our own version. It is now tempting to think that the main body of the message might reside in genomes (Marx, 1979; see also Hoch & Losick, 1997). Though the idea of genomic SETI (Davies, 2010) might seem naïve in view of random mutations, things are not so obvious. For example, a locus with a message might be exposed to purifying selection through coupling to essential genes, and there is even possible evidence for that (*ibid.*). Whatever the case, the ideogram does seem to provide a reference to genomes. Thus, complementary mirror-symmetrical bases of the short upper string (Fig. 9a) resemble Watson-Crick pairs; the four central bases TC|GA and the central axis therefore possibly represent the symbol of the genomic DNA itself. Flanking TATAT bases (Fig. 9b) might symbolize consensus sequence found in promoters of most genes. Coding sequences of genes are located between *Start*- and *Stop*-codons. Vice versa, nontranslated regions are found between *Stop*- and *Start*-codons of neighbor genes. Therefore the triplet string in Fig. 10c might symbolize intergenic regions, and may be interpreted as the address of the genomic message.

The privileged numeral system in the code might also be interpreted as an indication of a similar feature in genomes. It is often said that genomes store hereditary information in quaternary digital format. There are 24 possible numberings of DNA nucleotides with digits 0, 1, 2, 3. The ideogram seems to suggest the proper one: $T \equiv 0$, $C \equiv 1$, $G \equiv 2$, $A \equiv 3$. In this case the TCGA quadruplet (Fig. 9a), read in the distinguished direction, represents the natural sequence preceded by zero. Palindromic codons CCC and TCT (Fig. 10b) become a symbol of the quaternary digital symmetry $111_4$ and the radix of the corresponding system $010_4 = 4$, respectively. Translationally related AGC, or $321_4$, codons (Fig. 9b) possibly indicate positions in quaternary place-value notation, with higher orders coming first. The sum of digital triplets in the string TAG + TAA + AAA + ATG + ATG (Fig. 10c) equals to the number of nucleotides in the code $3000_4 = 192$. Besides, T as zero is opposed to the other three "digits" in the decomposed code (Fig. 6). Finally, each complementary base pair in DNA sums to 3, so the double helix looks numerically as $333..._4$, and the central AAA codon in Fig. 10c becomes the symbol of duplex DNA located between genes. Should this particular numbering have relation to the genomic message, if any, is a matter of further research.

It is worth mentioning that all genomes, despite their huge size and diversity, do possess a feature as universal as the genetic code itself. It is known as the second Chargaff's rule. In almost all genomes – from viral to human – the quantities of complementary nucleotides, dinucleotides and higher oligonucleotides up to the length of ~9 are balanced to a good precision within a *single* DNA strand (Okamura *et al.*, 2007). Unlike the first Chargaff's rule which quickly found its physicochemical basis, the second rule with its total orderliness still has no obvious explanation.

___________________________________________

**Appendix A. Molecular implementation of the genetic code**

Here we outline molecular workings behind the genetic code which explain why it stays unchanged for billions of years and, at the same time, might be readily modified artificially, e.g., for embedding a signal. For simplicity, we skip the details such as U instead of T in RNA, ATP energetics, wobble pairing, etc., that do not affect understanding of the main point (for details see, e.g., Alberts *et al.*, 2008).

The first type of molecules behind the genetic code is transfer RNAs (tRNAs). They deliver amino acids into ribosomes, where protein synthesis takes place. tRNAs are transcribed as final products from tRNA genes in genomes by RNA polymerase (Fig. A1a; for definiteness, the mechanism is shown for amino acid Ser and its TCC codon). With the length varying around 80 nucleotides, tRNA transcripts fold in a specific spatial configuration due to base-pairing between different sections of the same RNA strand, similar to as it occurs between two strands of DNA helix (Fig. A1b). At its opposite sides the folded tRNA molecule has an unpaired anticodon and the acceptor end to which amino acid is to be bound. tRNAs with differing anticodons specifying the same amino acid (remember the code is redundant) are identical in their overall configuration. tRNAs specifying distinct amino acids differ from each other in anticodons as well as other spots, so they have slightly different overall configurations. However, acceptor ends are identical in all tRNAs, so for tRNA itself it makes no difference which amino acid is bound to it, no matter which anticodon it has at the opposite side. The process of binding amino acids to tRNAs is performed by protein enzymes called aminoacyl-tRNA synthetases (aaRSs, Fig A1b, bottom). Normally, there are 20 types of aaRSs, one for each amino acid, and they themselves are translated from appropriate genes in genome. Each of these enzymes recognizes with great specificity both its cognate amino acid and all tRNAs specifying that amino acid; tRNAs are recognized primarily by their overall configuration, not exclusively by their anticodons (Fig. A1c). After binding and additional checking, aaRS releases tRNA charged with amino acid to be delivered to ribosome (Fig. A1d). In its turn, the ribosome does not care if tRNA carries an amino acid specified by its anticodon; it only checks if the anticodon of tRNA matches complementarily the current codon in messenger RNA (mRNA; Fig. A1e). If so, the amino acid is transferred from tRNA to the growing peptide chain and tRNA is released to be recycled. If codon and anticodon do not match, tRNA with its amino acid is dislodged from the ribosome to be used later until it matches codon on mRNA (even with this overshoot the bacterial ribosome manages to add ~20 amino acids per second to a peptide chain). The described mechanism results in relationships between mRNA codons and amino acids (Fig. A1f) which, collected together in any convenient form (one possibility is shown in Fig. 1a), constitute the genetic code.

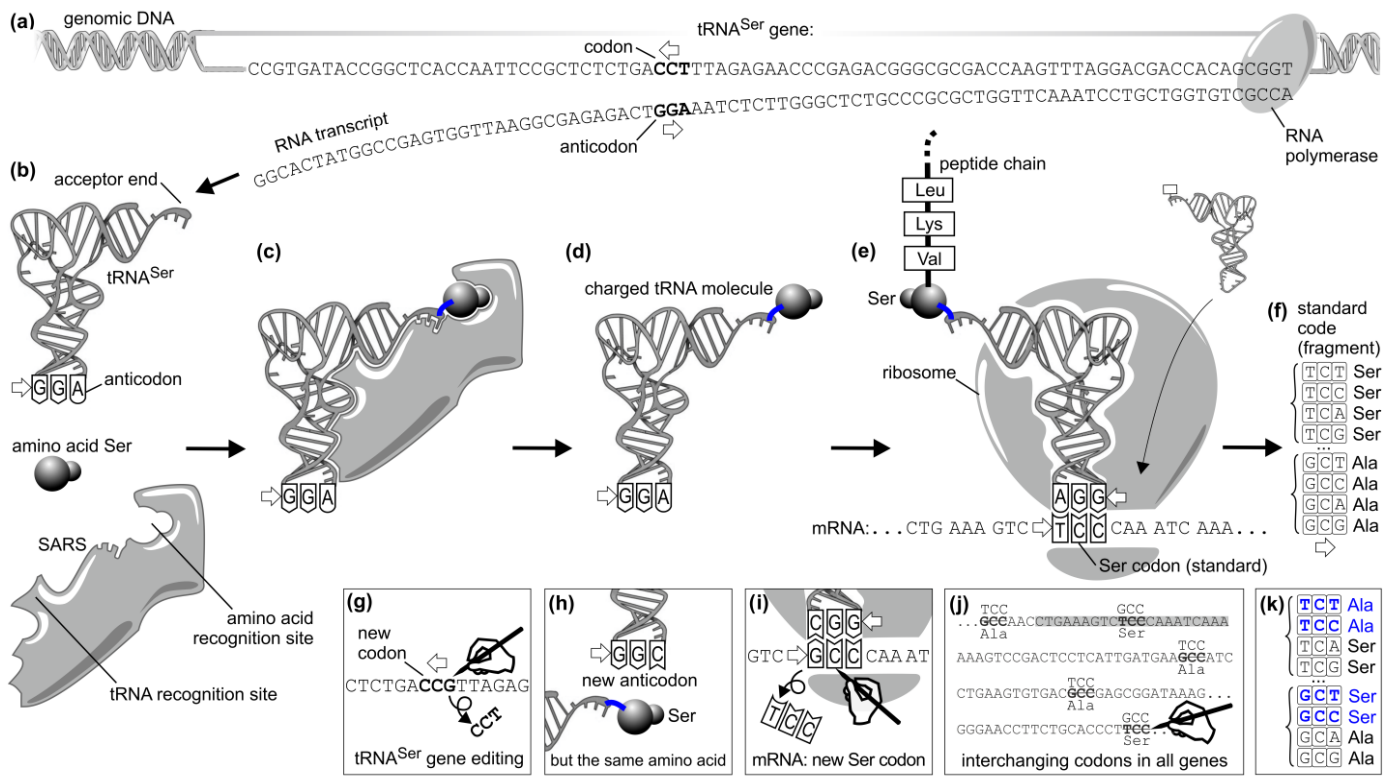

**Fig. A1.** Molecular mechanisms of the genetic code (shown for the case of amino acid serine) and a simple example of its artificial modification. The contour arrows indicate directionality of DNA and RNA strands as defined by orientation of their subunits (designated in biochemistry as 5′→3′ orientation; replication, transcription and translation occur only in that direction). **(a)** tRNA$^{Ser}$ gene (the gene of tRNA that specifies Ser in the standard code) is transcribed by RNA polymerase from genomic DNA. **(b)** The folded tRNA$^{Ser}$ molecule (top), serine molecule (middle) and seryl-tRNA synthetase (SARS, an aaRS cognate for amino acid serine; bottom). **(c)** SARS recognizes both serine and tRNA$^{Ser}$ and binds them together. **(d)** Ser-tRNA$^{Ser}$ released from SARS and ready to be delivered to ribosome. **(e)** The process of peptide synthesis at the ribosome (as an example, the mRNA with the gene fragment of the SARS itself is shown). **(f)** The resulting fragment of the genetic code (also shown is Ala group, which will be used in an example below). **(g)-(k)**. A simple way of genetic code modification. The shaded sequence in **(j)** corresponds to the region shown in **(e)**.

The key point in terms of changeability of the genetic code is that there is no direct chemical interaction between mRNA codons and amino acids at any stage. They interact via molecules of tRNA and aaRS both of which might be modified so that a codon is reassigned to another amino acid. As an example, Figures A1g-k show a simple way of changing the code where two amino acids – Ser and Ala – interchange two of their codons. It is known that in most organisms tRNA anticodons are not involved in recognition by aaRSs cognate for these amino acids (Giegé *et al*., 1998; the fact reflected in Fig. A1c with SARS not touching the anticodon). Therefore, the three nucleotides in tRNA$^{Ser}$ gene corresponding to anticodon might be replaced (Fig. A1g), in particular, to get GGC anticodon corresponding to GCC codon in mRNA, which normally codes Ala. (To get anticodon for a codon, or vice versa, one has to apply complementarity rule and reverse the resulting triplet, since complementary DNA/RNA strands have opposite directionalities). After that, SARS will still bind Ser to tRNA$^{Ser}$, even though it now has new GGC anticodon (Fig. A1h). If analogous procedure is performed with tRNA$^{Ala}$ genes to produce tRNA$^{Ala}$ with GGA anticodon, the genetic code would be modified: Ser and Ala would have interchanged some of their codons (actually, two codons, due to wobble pairing). However, the cell will not survive such surgery, since all coding genes in genome remain "written" with the previous code and after translation with the new code they all produce non- or at best semi-functional proteins, with Ala occasionally replaced by Ser and vice versa. To fix the new code in a cell lineage, one also has to change coding mRNAs appropriately to leave amino acid sequences of coded proteins unaltered (Fig. A1i). That would be automatically fulfilled if all coding genes are rewritten all over the genome so that TCC codons are replaced with GCC and vice versa (Fig. A1j); such operation is possible when genomes are even rewritten from scratch (Gibson *et al*., 2010). Now, amino acid sequences of proteins stay unaltered and a cell proliferates with the new genetic code (Fig. A1k).

It must be clear now why the genetic code is highly protected from casual modifications. If a mutation occurs in tRNA or aaRS leading to codon reassignment, all genes in genome remain written with the previous code, and a cell quickly goes off the scene without progeny. The chances that such mutation in tRNA/aaRS is accompanied by corresponding mutations in coding genes all over the genome resulting in unaltered proteins are vanishingly small, given that there are dozens of such codons in thousands of genes in a genome. Thus, the machinery of the genetic code experiences exceptionally strong purifying selection that keeps it unchanged over billions of years.

It should be reminded that in reality the process of intentional modification of the code is more complicated. For example, details of tRNA recognition by aaRSs vary depending on tRNA species and organism, and in some cases anticodon is involved, partially or entirely, in that process. However, this is avoidable, in principle, with appropriate methods of molecular engineering. Another issue is that modifications in the code that leave proteins unaltered still might affect the level of gene expression (Kudla *et al*., 2009). Therefore, additional measures might have to be taken to restore the expression pattern with the new genetic code. These are surmountable technical issues; the point is that there are no principal restrictions for changing the code artificially in any desired way. In effect, elaborate methods of modifying the overall tRNA configuration and/or aaRS recognition sites might allow not only interchanging two amino acids, but introducing new ones.

**Appendix B. Statistical test**

It is appropriate to ask if the presented patterns are merely an artifact of data fishing. To assess that, one might compare information volumes of the data set itself ($V_0$) and of the pattern ensemble within that set ($V_p$). The artifact of data fishing might then be defined as the case when $V_p \ll V_0$. As shown in *Appendix E*, the presented ensemble of patterns might be described with a system of Diophantine equations, where nucleon numbers of amino acids serve as unknowns. Given the set of canonical amino acids (the range of possible values for the unknowns), this system is completely defined: it has a single solution and that turns out to be the actual mapping of the code (this also implies that there are no more algebraically independent patterns of the same sort in the code). Hence, $V_p = V_0$, so the pattern ensemble employs informational capacity of the code entirely, showing that it represents a feature inherent to the code itself, rather than an artifact of data fishing.

One might ask then how likely such pattern ensemble is to appear in the genetic code by chance. Since this question implies that the current mapping of the code has been shaped by natural processes, it is more appropriate to ask how likely such pattern ensemble is to appear by chance under certain conditions reflecting presumable evolutionary pathways. We tested both versions of the null hypothesis ("the patterns are due to chance alone" and "the patterns are due to chance coupled with presumable evolutionary pathways"). The results are of the same order of magnitude; we describe only the version with presumable evolutionary pathways. Three such pathways reflecting predominant speculations on the code evolution were imposed on computer-generated codes in this test:

(1) Redundancy must be on average similar to that of the real code. This is thought to be due to the specifics of interaction between the ribosome, mRNA and tRNA (Novozhilov *et al.*, 2007). Besides, we took into account possible dependence of the probability for a codon family to stay whole or to be split on the type of its first two bases. This follows from the difference in thermostability between codon-anticodon pairs enriched with strong (G and C) bases and those enriched with weak (A and T) bases (Lagerkvist, 1978). For that, the probability for a family of four codons with leading strong doublets to specify a single amino acid was adopted to be 0.9, for those with weak doublets – 0.1, and for mixed doublets it was 0.5. Each of the 20 amino acids and *Stop* is recruited at least once; therefore codes with less than 21 generated blocks are discarded. After that blocks were populated randomly with amino acids and *Stop*.

(2) Reduced effect of mutations/mistranslations due to natural selection. The cost function for polar requirement was adopted from Freeland & Hurst (1998), taking into account transversion-transition and mistranslation biases (see also Novozhilov *et al.*, 2007). Only those codes were passed further which had cost function value smaller than $\varphi_0 + \sigma$, where $\varphi_0$ is the value for the universal code, and $\sigma$ is the standard deviation for all random codes filtered through the previous condition.

(3) Small departure from the cytoplasmic balance (see *Appendix D*). As argued by Downes & Richardson (2002), this balance might reflect evolutionary pathways optimizing the distribution of mass in proteins. With $C$ standing for all side chain nucleons in the code and $B$ for all nucleons in block residues, the value $\delta = (C - B)/(C + B)$ is distributed approximately normally with $\mu = 0.043$ and $\sigma = 0.024$ (under the first condition described above). Only those codes were considered which had $\delta$ in the range $0 \pm \sigma$, centered on the value of the standard code. As that range corresponds to codes with smaller ("early") amino acids predominating, this condition also reflects presumable history of the code expansion (Trifonov, 2000; Wong, 2005).

The random variable in question is the number of independent patterns of the same sort in a code. Obviously, the more such patterns are observed in a code, the less likely such observation is. Probably, a good approximation here would be a binomial distribution since, for example, a nucleon balance might be regarded as a Bernoulli trial: in a given arrangement the balance is either "on" or "off", where probability for "on" is much smaller than for "off". However, probabilities for balances in distinct arrangements might differ, especially under conditions imposed. Situation is even more complex with ideogram symmetries: symmetry is not just "on" or "off", it is also characterized by the length of a string and the number of nucleotide types involved. Therefore, we do not apply any approximations but use brute-force approach to find distributions for appropriately defined scores for the patterns. Proline was considered with one nucleon transferred from its side chain to its block (note that since the activation key is applied universally, the actual code and the code with the key applied are equivalent statistically).

*Nucleon balances.* Arithmetical patterns in the standard code are all of the same style: equality of nucleon sums + their distinctive decimal notation + at least one of the three transformations (except the decomposed case). The search for a random code with a few patterns of this sort turned out to be time-consuming, so the requirements were greatly simplified. Only nucleon equalities were considered, without requirement of distinctive notation in any numeral system. Presence of transformations was required only in Gamow's arrangement for codons with identical and unique bases, since transformations act there in the first place, not as companions of another sorting logic. Also for simplicity, only global patterns were considered; "local" features like the threefold balance in Fig. 5b were not checked.

Alternative codes might have balances in arrangements and combinations different from those in the real code. Contrary to as it might seem, there are not so many ways of arranging the code based on a straightforward logic with minimum arbitrariness. For example, along with Gamow's sorting, several other arrangements were proposed during early attempts to deduce the code theoretically (see Hayes, 1998). One of them is known as the "code without commas" (Crick *et al.*, 1957). However, unlike Gamow's sorting, this and other proposed arrangements do not allow "freezing" the code elements completely, leaving a large degree of arbitrariness. Ultimately, the following arrangements were considered in the test:
- divisions based on redundancy;
- divisions based on positions in codons (alternating all combinations such as S or W in the first position, R or Y in the second position, etc.);
- sortings based on nucleotide composition of codons (alternating all combinations of "freezing" conditions and division logic);
- arrangements based on decomposition of codons into bases (alternating all combinations of the four nucleotide sets).

Besides, the first two types might be arranged with full-size or contracted codons. The only possible balance of the peptide representation (*Appendix D*) was also checked. In total, 160 potential balances (of both chain-chain and block-chain types) were checked in all these arrangements. Precautions were made to ignore arithmetical dependencies, as for certain code versions some balances are trivially fulfilled if few others occur. A simple scoring scheme was adopted: the score of a code is the number of algebraically independent nucleon equalities it happens to possess in all arrangements. In this scheme the simplified version of arithmetical patterns in the standard code has the score 7. Computer estimation shows that probability for a code to have the score not less than 7 by chance under imposed conditions is $p_1 = 1.5 \times 10^{-8}$ (Fig. B1a).

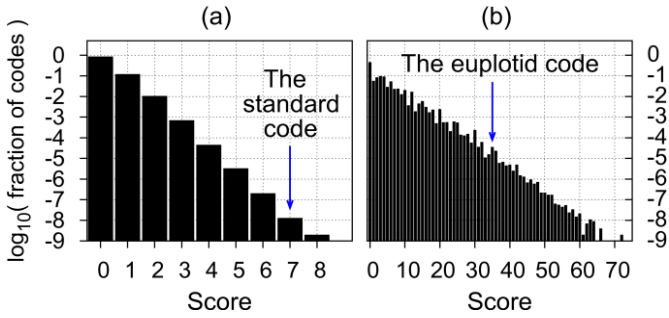

**Fig. B1.** Distribution of variant codes by their scores for **(a)** nucleon equalities and **(b)** ideogram symmetries. The size of the sample in both cases is one billion codes.

*Ideogram symmetries.* An ideogram might be built for each variant code in the same way as shown in Fig. 7 (however, no requirement is imposed for whole and split families to be linked with any transformation). There are a few more conceivable ways to build an ideogram using contracted codon series (ideograms based on full-size codons suffer with ambiguities). For example, nucleon and redundancy numbers might be arranged in the same direction, rather than antisymmetrically. Another way is to divide the code by positions in codons (e.g., R or Y in the first position; though these ideograms are simpler as two of their four upper strings are always binary, whereas in ideograms based on redundancy all strings are, in general, quaternary). In total, 9 ideogram versions were built for each code and checked for symmetries. Namely, each of the four strings was checked for $\mathcal{M}$, $\mathcal{M} + I$, $\mathcal{T}$, $\mathcal{T} + I$, where $\mathcal{M}$ and $\mathcal{T}$ stand for mirror and translation symmetries and $I$ denotes pair inversions of all three types. For each symmetry a string of length $L$ gets the score $L/2$, if it contains only two types of bases (or if the symmetry holds only in binary representation RY, SW or KM), and $L$, if it contains three or all four types of bases. Only whole-string symmetries were considered (in this case multiple symmetries organizing different parts of a string such as in Fig. 9b are not detected; the whole string in Fig. 9b, however, is mirror symmetrical in KM representation). For each ambiguous position (two neighboring series with equal nucleon numbers) the penalty $L/3$ was introduced. Semantical symmetries and balances of translated amino acids were not checked. Finally, if at least one of the four strings has none of the symmetries, the score is divided by 2. The euplotid code has the score 35 in this scheme: 8 for $\mathcal{M} + I_{(T \leftrightarrow A, C \leftrightarrow G)}$ and 4 for $\mathcal{T}_{RY}$ in the upper short string, 4 for $\mathcal{M}_{RY}$ in the center short string, 8 for $\mathcal{M}_{KM}$ in the upper long string, 16 for $\mathcal{M}$ in the center long string, penalty $-16/3 \approx -5$ for Lys and Gln (though in this case their interchange affects neither $\mathcal{M}_{KM}$ in the upper string, nor $\mathcal{M}$ in the center one). Computer estimation shows that probability for a code to have the score not less than 35 by chance under imposed conditions is $p_2 = 9.4 \times 10^{-5}$ (Fig. B1b).

We also checked transformations in Rumer's bisections of generated codes, since these transformations served as the guiding principle for signal extraction in the real code. Under the conditions imposed, probability for a random code to have equal numbers of whole and split families which are furthermore linked with any of the three possible transformations was found to be $4.6 \times 10^{-2}$. Given that one transformation takes place, the other two might be distributed among codons in the ratios 8:0 ($p = 0.125$), 4:4 ($p = 0.375$), or 2:6 ($p = 0.5$). For the real code this ratio is 4:4 (see Fig. 2a), so finally $p_3 = 1.7 \times 10^{-2}$.

As suggested by a separate computational study, mutual influence of the three types of patterns is negligible, so the total probability for a (very simplified) signal to occur by chance in a single code under imposed conditions is $p_1 p_2 p_3 = 2.4 \times 10^{-14}$. Since the redundancy-symmetric code is not even needed to be found in nature to reveal the ideogram, the final *P*-value will not differ much from that value.

This result gives probabilities for the specific type of patterns – nucleon equalities, ideogram symmetries and transformations. However, testing the hypothesis of an intelligent signal should take into account patterns of other sorts as well, as long as they meet the requirements outlined in *Introduction*. After analysis of the literature on the genetic code our opinion is still that nucleon and redundancy numbers are the best candidates for "ostensive numerals". We accept though that there could be other possibilities and that the obtained *P*-value should be regarded as a rough approximation (keep in mind simplifications in the test as well). But admittedly, there are just not enough candidates for "ostensive numerals" and corresponding (algebraically defined) pattern ensembles to compensate for the small *P*-value obtained and to raise it close to the significance level.

**Appendix C. Digital symmetries of positional numeral systems**

The digital symmetry described in the main text for the decimal system is related to a divisibility criterion that might be used to effectively perform checksums. Consider the number 27014319417 as an example. Triplet reading frame splits this number into digital triplets 270, 143, 194, 170 (any of the three reading frames might be chosen; zeros are added at flanks to form complete triplets). The sum of these triplets equals to 777. Its distinctive notation indicates that the original number is divisible by *037*. In four-digit numbers that appear during summations thousand's digits are transferred to unit's digits. If notation of the resulting sum is not distinctive, add or subtract *037* once. Subsequent distinctive notation will confirm the divisibility of the original number by *037* while its absence will disprove it. Thus, the other two frames for the exemplary number yield:

002 + 701 + 431 + 941 + 700 = *2775* → 002 + 775 = 777;
027 + 014 + 319 + 417 = 777.

This criterion applies to numbers of any length and requires a register with only three positions. Moving along a linear notation, such register adds digital triplets together and transfers thousand's digits to unit's digits.

The same triplet digital symmetry and related divisibility criterion exist in all numeral systems with radix $q$ that meets the requirement $(q - 1)/3 = Integer$. The symmetry-related prime number in those systems is found as $111_q/3$. Thus, the feature exists in the quaternary system ($q = 4$) with prime number 7 (*013*$_4$), septenary system ($q = 7$) with prime number 19 (*025*$_7$), decimal system ($q = 10$) with prime number *037*, the system with $q = 13$ and prime number 61 (*049*$_{13}$), and so on. The digital symmetry of the quaternary system is shown in Fig. C1.

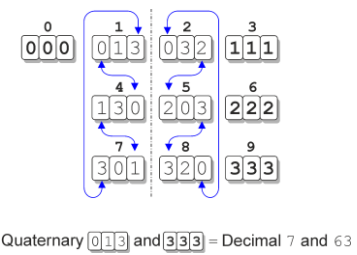

**Fig. C1.** Similar to the decimal system, the quaternary system also displays symmetry of digital triplets, where 7 (*013*$_4$) acts instead of *037*.

**Appendix D. The cytoplasmic balance**

Fig. D1 represents the entire genetic code as a peptide. Each amino acid is inserted into this peptide as many times as it appears in the standard code. Amino acid block residues make up the peptide backbone. The resulting polymer is 61 amino acids long. If its N- and C-termini are eliminated by closing the peptide into a ring, its backbone and side chains appear precisely balanced. Notably, this feature is common to natural proteins: their mass is distributed approximately equally between peptide backbone and side chains (Downes & Richardson, 2002). This also automatically implies that frequencies of amino acids in natural proteins correlate with their abundance in the genetic code (see data in Gilis *et al.*, 2001).

Not only the activation key is discarded in this balance, but amino acid molecules are considered as they appear in cytoplasmic environment (where side chains of some of them are ionized). For these reasons the balance shown in Fig. D1 is referred to as natural or cytoplasmic. Nevertheless, unusual peptide form (though circular peptides do occur rarely in nature, see Conlan *et al.*, 2010) and distinction between amino acid blocks and chains suggest that the cytoplasmic balance and the "virtual" balances shown in the main text are likely to be related phenomena. Possibly, this balance is intended to validate the artificial nature of the activation key, showing that only actual proline can maintain patterns in natural environment. This balance was found by Downes & Richardson (2002) from biological aspect. Simultaneously, Kashkarov *et al*. (2002) found it with a formal arithmetical approach.

**Appendix E. Algebraic representation of the signal**

Here we describe a possible way the signal-harboring mapping might have been obtained. As initial data, one has a set of 64 codons and another set of 20 canonical amino acids plus *Stop*. Suppose, the mapping between those two sets is unknown and it has to be deduced from the given pattern ensemble of the signal. There are ~$10^{83}$ possible mappings between the two sets, provided that each element from the second set is represented at least once. Knowing the ideogram (without knowing nucleon numbers mapped to individual codons) is equivalent to knowing the block structure of the code. From this follows the first portion of equations $ggt = ggc = gga = ggg = ggn$, $ttt = ttc = tty$, etc., where codons are used to denote variables – unknown nucleon numbers of amino acid side chains. Thus, the number of elements in the first set is essentially reduced from 64 to 24. But there are still ~$10^{30}$ possible mappings left. Now one might write down the nucleon sums from Figs. 5-8 and 10 (leaving out algebraically dependent parts and standard block sums, as we are provided with the set of canonical amino acids; in case of projecting the patterns *Stop* might be preliminarily assigned to certain codons to make things easier with the block sums):

$ggn + gcn + tcn + ccn + gtn + acn + ctn + cgn = 333$ (Fig. 7b);
$tgy + tga + ath + tar + agy + ttr + aay + gay + car + aar + gar + cay + tty + agr + tay + atg + tgg = 111 + 999$ (Fig. 7b);
$tty + ttr + tcn + tay + tar + tgy + tga + tgg + ctn + ccn + cay + car + cgn = 814$ (Fig. 8a);
$tty + ttr + tcn + tay + tar + tgy + tga + tgg + gtn + gcn + gay + gar + ggn = 654$ (Fig. 8b);
$tty + ttr + ctn + ath + atg + gtn + tgy + tga + tgg + cgn + agy + agr + ggn = 789$ (Fig. 8b);
$tty + aar + ath + tcn + cay + 2gcn + ctn + tgy + tga + gay + atg + car + agy = 703$ (Fig. 5a);
$ggn + ccn + ctn + 2acn + tay + tcn + 2gtn + 2cgn + agy + tar + gay = 703$ (Fig. 5a);
$tty + 2ttr + 3ccn + 2ctn + ath + gtn + 2tcn + acn + gcn + tay + tgy + cay + cgn = 999$ (Fig. 5b);

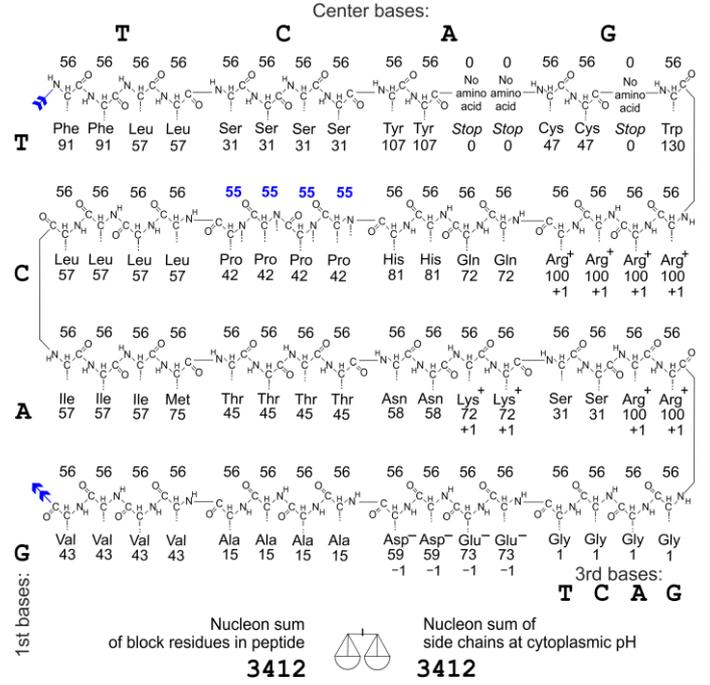

**Fig. D1.** Amino acids of the standard genetic code in the form of a circular peptide (sequence order does not matter). The peptide is formed by aggregating standard blocks of amino acids into polymer backbone. Formation of each peptide bond releases a water molecule reducing each amino acid block to 56 nucleons (55 in proline). Asp and Glu lose one proton each from their side chains at cytoplasmic pH, while Arg and Lys gain one proton each (denoted with –1 and +1, respectively). Other amino acids are predominantly neutral in cytoplasmic environment (Alberts *et al.*, 2008). As a result, nucleon sum of the peptide backbone is exactly equal to that of all its side chains.

$2aay + aar + tar + car + gar = 333$ (Fig. 5b);
$3ggn + tgg + cgn + agr = 333$ (Fig. 5b);
$ath + acn + agr + gtn + gcn + gar = 333$ (Fig. 5b);
$tty + 2ctn + 2tcn + ccn + 2aay + tar + ath + car + acn + 2ggn + tgg + gtn + cgn + gcn = 888$ (Fig. 5c);
$5tty + 4ttr + 5ctn + 4ath + atg + 5gtn + 5tcn + ccn + acn + gcn + 3tay + 2tar + cay + aay + gay + 3tgy + tga + tgg + cgn + agy + ggn = 666 + 999 \times 2$ (Fig. 6b);
$2tar + aar + 2atg = 222$ (Fig. 10d);
$agy + 2aar + tgh = 222$ (Fig. 10e).

The cytoplasmic balance is not accounted here as it has no algebraic connection to this system due to the activation key. There are also additional inequalities provided by the ideogram (Fig. 7a):

$ggn \leq gcn \leq tcn \leq ccn \leq gtn \leq acn \leq ctn \leq cgn$;
$tgh \leq ath$;
$tar \leq agy \leq ttr \leq aay \leq gay \leq car \leq aar \leq gar \leq cay \leq tty \leq agr \leq tay$;
$atg \leq tgg$.

Finally, $tgh = tgy$ to account for two code versions. In total, there are 26 unknowns, 16 equations and 20 inequalities. Generally, such systems of Diophantine equations and inequalities have multiple solutions. Since we are interested here in obtaining the mapping of the code given the patterns and the fixed set of canonical amino acids plus *Stop*, the solution is to be searched over the fragmentary domain of integers and zero {0, 1, 15, 31, 41, 43, 45, 47, 57, 58, 59, 72, 73, 75, 81, 91, 100, 107, 130}. In this case analysis of the system with any computer algebra system capable of dealing with Diophantine expressions shows that this system has a single solution coinciding with the actual mapping of nucleon numbers onto codons: $tty = 91$, $ggn = 1$, $tga = 0$, $ath = 57$, etc. That still leaves us with several mappings for amino acids though, since two of the roots – 57 and 72 – represent two amino acids each. This ambiguity is eliminated when the patterns within side chains (Figs. 7b and 8a) are also taken into account. After that the actual mapping of the code is deduced unambiguously from the algebraic system of the patterns. In fact, analysis shows that unambiguous solution is achieved even if the restriction of fragmentary domain is applied only to some of the unknowns. In another approach (*sh*Cherbak, 2003) unambiguous solution is achieved only with few assumptions about the amino acid set.

### Acknowledgments

The study was partially financed by the Ministry of Education and Science of the Republic of Kazakhstan. The research was appreciably promoted by Professor Bakytzhan T. Zhumagulov from the National Engineering Academy of the Republic of Kazakhstan. Part of the research was made during V.I.S.' stay at Max-Planck-Institut für biophysikalische Chemie (Göttingen, Germany) on kind invitation of Professor Manfred Eigen. V.I.S. expresses special thanks to Ruthild Winkler-Oswatitsch for her valuable help and care. M.A.M. acknowledges the support by the administration of Fesenkov Astrophysical Institute. The authors are grateful to Professor Paul C.W. Davies, Felix P. Filatov, Vladimir V. Kashkarov, Artem S. Novozhilov, Denis V. Tulinov, Artem N. Yermilov and Denis V. Yurin for objective criticism and fruitful discussions of the manuscript. We deeply appreciate Icarus Editorial Staff for organizing the peer review and the three reviewers for their comments which led to the improvement of the manuscript.

### Authors' contributions

V.I.S. conceived of and performed the research, developed graphic arts. V.I.S. and M.A.M. analyzed data, introduced interpretation of the activation key, outlined structure of the paper. M.A.M. performed statistical test and algebraic analysis, wrote the manuscript.